\documentclass[twoside,12pt]{article}
\usepackage{CJK}
\usepackage{indentfirst}
\usepackage{bm}
\usepackage{epsfig}
\usepackage{amssymb}
\usepackage{amsmath}

\begin{document}
	
	\begin{CJK}{GBK}{song}
		
		\thispagestyle{empty} \vspace*{0.8cm}\hbox
		to\textwidth{\vbox{\hfill\huge\sf Commun. Theor. Phys.\hfill}}
		\par\noindent\rule[3mm]{\textwidth}{0.2pt}\hspace*{-\textwidth}\noindent
		\rule[2.5mm]{\textwidth}{0.2pt}
		
		
		\begin{center}
			\LARGE\bf Escape rate of an active Brownian particle in a rough potential
		\end{center}
		
		\footnotetext{\hspace*{-.45cm}\footnotesize $^\dag$Corresponding author: tuzc@bnu.edu.cn }

		\begin{center}
			\rm Yating Wang$^{\rm a)}$, and Z. C. Tu$^{\rm a)\dagger}$
		\end{center}
		
		\begin{center}
			\begin{footnotesize} \sl
				${}^{\rm a)}$Department of Physics, Beijing Normal University, Beijing100875, China \\
				
			\end{footnotesize}
		\end{center}
		
		\begin{center}
			\footnotesize (Received XXXX; revised manuscript received XXXX)
			
		\end{center}
		
		\vspace*{2mm}
		
		\begin{center}
			\begin{minipage}{15.5cm}
				\parindent 20pt\footnotesize
				We discuss escape problem with the consideration of both the activity of particles and the roughness of potentials. we derive analytic expressions for the escape rate of a Brownian particle (ABP) in two types of rough potentials by employing the effective equilibrium approach and the Zwanzig method. We find that activity enhances the escape rate, but both the oscillating perturbation and the random amplitude hinder escaping. 
			\end{minipage}
		\end{center}
		
		\begin{center}
			\begin{minipage}{15.5cm}
				\begin{minipage}[t]{2.3cm}{\bf Keywords:}\end{minipage}
				\begin{minipage}[t]{13.1cm}
					escape rate, active Brownian partiale, rough potential, effective potential.
				\end{minipage}\par\vglue8pt
				
			\end{minipage}
		\end{center}
		
		\section{\label{sec:level1}Introduction}
		
		Escape problem has attracted much attention of researchers in various fields$^{[1-10]}$. The Arrhenius formula indicates that the rate of chemical reaction depends exponentially on inverse temperature$^{[3,4]}$. Kramers presented the transition state method for calculating the rate of chemical reactions by considering a Brownian particle escaping over a potential barrier$^{[5]}$. Subsequent studies on escape rate are summarized in Ref. $[10]$. All of the above studies merely involve passive particles. The research theme has been transferred to active particles with self-propulsion in recent years$^{[11-21]}$. Active systems are intrinsically non-equilibrium since the detailed balance is broken. An effective equilibrium method has been developed to investigate active Brownian particles$^{[22-25]}$. By using this method, Sharma $et$ $al$. discussed an escape problem of active particles in a smooth potential$^{[26]}$. They found that introducing activity increases the escape rate.

		The escape problem in the researches mentioned above is simplified as a Brownian particle climbing over a smooth potential barrier. However, the potential is not always smooth in reality. Interface area scans of proteins imply that the protein surface is not smooth$^{[27,28]}$. Hierarchical arrangement of the conformational substrates in myoglobin indicates that the potential surface might be rough$^{[29]}$. In addition, the inside of the cell is quite crowed. Thus, diffusion of substance in the cell may not be regarded as Brownian motion in smooth potential. In the biochemical point of view, it is valuable to consider the influence of the roughness of potential to diffusion behaviors. The study of diffusion in rough potential offers insight into fields from transport process in disordered media$^{[30,31]}$ to protein folding$^{[32,33]}$ and glassy systems$^{[34,35]}$. Zwanzig dealt with diffusion in a rough potential and found that the roughness slows down the diffusion at low temperatures$^{[36]}$. Roughness-enhanced transport was also observed in ratchet systems$^{[37-39]}$. Hu $et$ $al$ discussed diffusion crossing over a barrier in a random rough metastable potential$^{[40]}$. By using numerical simulations, they demonstrate that a decrease in the steady escape rate in with the increase of rough intensity. Activity of particles was not considered in these works. 
		
		There are a large number of active substances, biochemical reactions, and transport of substances in organism. Therefore, it is of practical significance to discuss escape problem with the consideration of both the activity of particles and the roughness of potentials. In this work, we calculate the escape rate of an active Brownian particle (ABP) in rough potentials by using the effective equilibrium approach$^{[22-26]}$ and the Zwanzig method$^{[36]}$. The rest of this paper is organized as follows: In section $2$, we briefly introduce the effective equilibrium approach. In section $3$, we discuss the escape problems of ABPs in rough potentials with oscillating perturbation or random amplitude. We derive the effective rough potentials following the effective equilibrium approach. Then we analytically calculate the escape rates of ABPs in the effective rough potentials. We find that activity enhances the escape rate, but both the oscillating perturbation and the random amplitude hinder escaping. The last section is a brief summary.
		
		\section{\label{sec:level2} Effective equilibrium  approach}
		In this section, we briefly revisit the main ideas of effective equilibrium approach$^{[22-26]}$. 
		
		The motion of the ABP can be described by the following overdamped Langevin equations 
		\begin{equation}\label{eq:1}
		\dot{\mathbf{r}} = v_{0}\mathbf{n} + \gamma^{-1} \mathbf{F} +\bm{\xi}(t),
		\end{equation}
		\begin{equation}\label{eq:2}
		\dot{\mathbf{n}} = \bm{\eta}(t) \times \mathbf{n},
		\end{equation}
		where $\gamma$ is the friction coefficient and $\mathbf{F}(t)$ is force on the ABP. $\mathbf{r}$ 
		represents position of the particle. The particle is self-propelling with constant speed $v_0$ along orientations $\mathbf{n}$. The dot ``$\cdot$" above a character represents the derivative with respect to time $t$. The stochastic vectors $\bm{\xi}(t)$ and $\bm{\eta}(t)$ are white noise with correlations $\left\langle \bm{\xi}(t) \bm{\xi}(t')\right\rangle=2 D_t\mathbf{I}\delta(t-t')$ and $\left\langle \bm{\eta}(t) \bm{\eta} (t')\right \rangle =2 D_r \mathbf{I}\delta(t-t')$, where $D_t$ and $D_r$ are the translational and rotational diffusion coefficients, respectively. $\mathbf{I}$ is the unit tensor.
		
		We obtain $\left\langle \mathbf{n}(t) \right\rangle=0$ and $\left \langle \mathbf{n}(t)\mathbf{n}(t')\right \rangle =(1/3)\mathbf{I}e^{-2D_r|t-t'|}$ from Eq.~(\ref{eq:2}). Substitute them into Eq.~(\ref{eq:1}), we derive
		\begin{equation}\label{eq:3}
		\dot{\mathbf{r}} = \gamma^{-1} \mathbf{F}+\bm{\chi}(t),
		\end{equation}
		where $\left\langle\bm{\chi}(t)\right\rangle=0$ and $\left \langle \bm{\chi}(t)\bm{\chi}(t')\right \rangle =2 D_t\mathbf{I}\delta(t-t')+(v^2_0/3)\mathbf{I}e^{-2D_r|t-t'|}$.
		
		A stochastic process with color-noise in Eq.~(\ref{eq:3}) is non-Markovian. It is impossible to derive an exact Fokker-Planck equation for the time evolution of the probability distribution. Nevertheless, using the Fox approximate method$^{[41,42]}$, we may derive an approximate Fokker-Planck equation  
		\begin{equation}\label{eq:4}
		\frac{\partial\bm{\phi}(\mathbf{r},t)}{\partial t} = -\nabla \cdot \mathbf {J}(\bm{r},t),
		\end{equation} 
		where $\bm{\phi}(\mathbf{r},t)$ is the probability distribution. The current $\mathbf{J}(\mathbf{r},t)$ is expressed as
		\begin{equation}\label{eq:5} 
		\mathbf{J}(\mathbf{r},t)=-D_t D(\mathbf{r}) \left[\nabla -\beta\mathbf{F}^{\rm eff}(\mathbf{r})\right] \bm{\phi}(\mathbf{r},t),
		\end{equation}
		where $\mathbf{F} ^{\rm eff}(\mathbf{r})$ represents the effective force on the particle. $\beta=(k_{\rm B}T)^{-1}$, in which $k_{\rm B}$ is the Boltzmann constant and $T$ is the temperature. The dimensionless effective diffusion coefficient 
		$D(\mathbf{r})=1+D_a/(1-\tau\nabla \cdot  \beta\mathbf{F}(\mathbf{r}))$,
		where $\tau=D_t/(2D_r)$. The activity parameter $D_a=v^2_0/(6D_r D_t)$. The effective force is given by 
		\begin{equation}\label{eq:7}
		\mathbf{F}^{\rm eff}(\mathbf{r})=\frac{1}{D(\mathbf{r})}\left[\mathbf{F}(\mathbf{r})-\beta\nabla D(\mathbf{r})\right]
		\end{equation}
		
		\section{\label{sec:level3}Escape rate of ABP in rough potentials} 
		In this section, we will deduce the effective rough potential and escape rate of ABP in rough potentials. For simplicity, we only consider the case that the bare force depends merely on a one-dimension potential $V=V(x)$. In this case, $\mathbf{F}=-V'(x)\mathbf{i}$, where $\mathbf{i}$ is the unit vector of $x$-coordinate. The prime ``$\prime$" on the top right of a character represents the derivative with respect to position $x$. From Eq.~(\ref{eq:7}) we can obtain the effective potential 
		\begin{equation}\label{eq:8}
		\beta V^{\rm eff}(x)=\ln D(x)+ \int_{0}^{x}dy\frac{\beta V'(y)}{D(y)}
		\end{equation}
		with
		\begin{equation}\label{eq:8b}
		D(x)=1+\frac{D_a}{1+\tau \beta V''(x)}.
		\end{equation}
		
		Now, let us considering a rough potential   
		\begin{equation}\label{eq:9}
		\beta V(x)=\frac{1}{2}\kappa_0 x^2 -\alpha x^3 + \varepsilon V_1(x),
		\end{equation}
		where $\kappa_0$ and $\alpha$ are positive constants. The first two terms in Eq.~(\ref{eq:9}) provide a smooth background with barrier. The last term in Eq.~(\ref{eq:9}) is the superposed random or oscillating perturbation. The amplitude $\varepsilon$ is assumed to be small, which represents a measure of the ``roughness" of the potential. 
		
		Now, we look for the effective rough potential $\beta V^{{\rm eff}}(x)$ corresponding to Eq.~(\ref{eq:9}) from Eq.~(\ref{eq:8}). Assuming $\kappa_0\tau\ll 1$ and keeping the terms up to the linear order of $\kappa_0\tau$ and $\varepsilon$, we obtain the effective rough potential 
		\begin{eqnarray}\label{eq:11}
		\beta V^{{\rm eff}}(x)\approx\frac{1}{2}\kappa_a x^2 -\alpha' x^3+g(x)+\frac{\varepsilon V_1(x)}{1+D_a},
		\end{eqnarray}
		where
		\begin{equation}\label{eq:12}
		\kappa_a=\kappa_0\left[\frac{1}{1+D_a}+\frac{D_a \kappa_0 \tau}{(1+D_a)^2}\right],
		\end{equation}
		\begin{equation}\label{eq:13}
		\alpha'=\alpha\left[\frac{1}{1+D_a}+\frac{3D_a \kappa_0 \tau}{(1+D_a)^2}\right],
		\end{equation}
		\begin{equation}\label{eq:14}
		g(x)=\frac{6D_a\alpha \tau }{1+D_a}x+\frac{9D_a \alpha^2 \tau}{2(1+D_a)^2}{x}^4.
		\end{equation}
		The above three equations and the first three terms in Eq.~(\ref{eq:11}) have been derived in Ref. $[26]$. 
		
		The bare and effective rough potentials are schematically depicted in Fig. 1. $x_a$ and $x_b$ correspond to the minimum and maximum of the potential, respectively. $x_c$ is a point on the right of $x_b$. Passing $x_c$, the particle will not return. In stationary state, the current~(\ref{eq:5}) can be rewritten as
		\begin{equation}\label{eq:15}
		J_{\rm act}^{\rm rou}=-D_t D(x)e^{- \beta V^{\rm eff}(x)}\frac{d}{dx}\left[e^{\beta V^{\rm eff}(x)}\phi(x)\right].
		\end{equation}
		Following Kramers's approach$^{[5,43]}$, we obtain the inverse of escape rate of ABP:
		\begin{equation}\label{eq:16}
		\frac{1}{r_{\rm act}^{\rm rou}}=\int_{x_1}^{x_2}dye^{-\beta V^{\rm eff}(y)}\int_{x_a}^{x_c}dz\frac{1}{D_t D(z)}e^{\beta V^{\rm eff}(z)},
		\end{equation}
		where $x_1 \leq x_a \leq x_2 \leq x_b$. The detailed derivation of this equation is shown in Appendix A.
		
		Considering the rough character of the potential, we use the Zwanzig method$^{[36]}$ to simplify Eq.~(\ref{eq:16}). The rough potential~(\ref{eq:11}) may be decomposed into two parts. One is the smooth skeleton
		\begin{eqnarray}\label{eq:11a}
		\beta V_0^{{\rm eff}}(x)=\frac{1}{2}\kappa_a x^2 -\alpha' x^3+g(x),
		\end{eqnarray}
		the other is the rough perturbation 
		\begin{eqnarray}\label{eq:11b}
		\beta V_1^{{\rm eff}}(x)=\frac{\varepsilon V_1(x)}{1+D_a}.
		\end{eqnarray}
		Since $V_1^{{\rm eff}}(x)$ varies quickly with $x$, we consider its average effect on escape rate in Eq.~(\ref{eq:16}). Define $\psi^+(x)$ and $\psi^-(x)$ such that 
		\begin{equation}\label{eq:111}
		e^{\psi^{\pm }(x)}=\left\langle e^{\pm \beta V_1^{\rm eff}(x)}\right\rangle,
		\end{equation}
		where $\left\langle ~ \right\rangle$ denotes the spatial average during a small interval $(x-\Delta/2,x+\Delta/2)$. Then Eq.~(\ref{eq:16}) is transformed into 
		\begin{equation}\label{eq:20}
		\frac{1}{r_{\rm act}^{\rm rou}}
		=\int_{x_a}^{x_c}dye^{-\beta V_0^{\rm eff}(y)}e^{\psi^{- }(y)}\int_{x_a}^{x_c}dz\frac{e^{\beta V_0^{\rm eff}(z)}e^{\psi^{+ }(z)}}{D_t D(z)}.
		\end{equation}
		Next we discuss the spacial situation that $\psi^\pm (x)$ happens to be independent of $x$. In this case, the above equation is transformed into 
		\begin{equation}\label{eq:201}
		\frac{1}{r_{\rm act}^{\rm rou}}
		=e^{\psi^{- }}e^{\psi^{+ }}\int_{x_a}^{x_c}dye^{-\beta V_0^{\rm eff}(y)}\int_{x_a}^{x_c}dz\frac{e^{\beta V_0^{\rm eff}(z)}}{D_t D(z)}.
		\end{equation} 
		By using the saddle-point approximation and considering $\kappa_0\tau$ is small, we derive the escape rate 
		\begin{equation}\label{eq:23}
		r_{\rm act}^{\rm rou}=\frac{ D_t(1+D_a)\sqrt{\left|\kappa_a\kappa_b\right|}e^{-(\beta E_b-\frac{D_a\kappa_0\tau}{1+D_a})}}{2\pi e^{\psi^-}e^{\psi^+}},
		\end{equation}
		where 
		\begin{equation}\label{eq:21}
		\kappa_b=\kappa_0\left[-\frac{1}{1+D_a}+\frac{D_a\kappa_0\tau}{(1+D_a)^2}\right],
		\end{equation}
		and 
		\begin{equation}\label{eq:22}
		\beta E_b=\frac{\kappa_0^3}{54\alpha^2(1+D_a)}+\frac{2D_a\kappa_0\tau}{1+D_a}.
		\end{equation}
		The detailed derivation of Eq~(\ref{eq:23}) is displayed in Appendix B. 
		
		For a passive Brownian particle moving in a smooth potential, Eq.~(\ref{eq:23}) is degenerated into
		\begin{eqnarray}\label{eq:24}
		r_{\rm pass}=\frac{ D_t \kappa_0}{2\pi}e^{-\beta E_0},
		\end{eqnarray}
		where $\beta E_0=\kappa_0^3/(54\alpha^2)$. 
		This is exactly the Kramers rate for the passive particle escaping from a smooth barrier$^{[5]}$. The escape rate of ABP in rough potential may be further expressed as 
		\begin{eqnarray}\label{eq:241}
		r_{\rm act}^{\rm rou}
		=r_{\rm pass}e^{\frac{D_a\left(\beta E_0-\kappa_0\tau\right)}{1+D_a}}[e^{\psi^-}e^{\psi^+}]^{-1}.
		\end{eqnarray}
		Obviously, the above equation implies the escape rate 
		\begin{equation}\label{eq:291}
		r_{\rm act}=r_{\rm pass}e^{\frac{D_a(\beta E_0-\kappa_0\tau)}{1+D_a}}
		\end{equation}
		for APB in a smooth potential$^{[26]}$ since $\psi^ +=\psi^ -=1$ for the smooth potential. Then, Eq.~(\ref{eq:24}) can be further expressed as
		\begin{equation}\label{eq:29}
		r_{\rm act}^{\rm rou}=r_{\rm act}[e^{\psi^-}e^{\psi^+}]^{-1}.
		\end{equation}
		
		\subsection{\label{sec:level} Oscillating perturbation of rough potential}

		We consider the oscillating perturbation, $V_1(x)={\rm sin}(qx)$ where $q\gg\sqrt{\kappa_0}$. Using Eq.~(\ref{eq:11}), the effective rough potential may be expressed as
		\begin{eqnarray}\label{eq:26}
		\beta V^{{\rm eff}}(x)\approx\frac{1}{2}\kappa_a x^2 -\alpha' x^3+g(x)+\frac{\varepsilon{\rm sin}(qx)}{1+D_a}.
		\end{eqnarray}
		
		In Fig.~\ref{fig1}, we plot the effective potential for different values of activity parameter $D_a$. We find that the effective barrier decreases with the increase of the activity parameter. Thus, the introduction of activity lowers the effective barrier height so that the particle easily escapes the barrier.
		
		\begin{figure}\label{fig1}
			\centering
			\includegraphics[width=7cm]{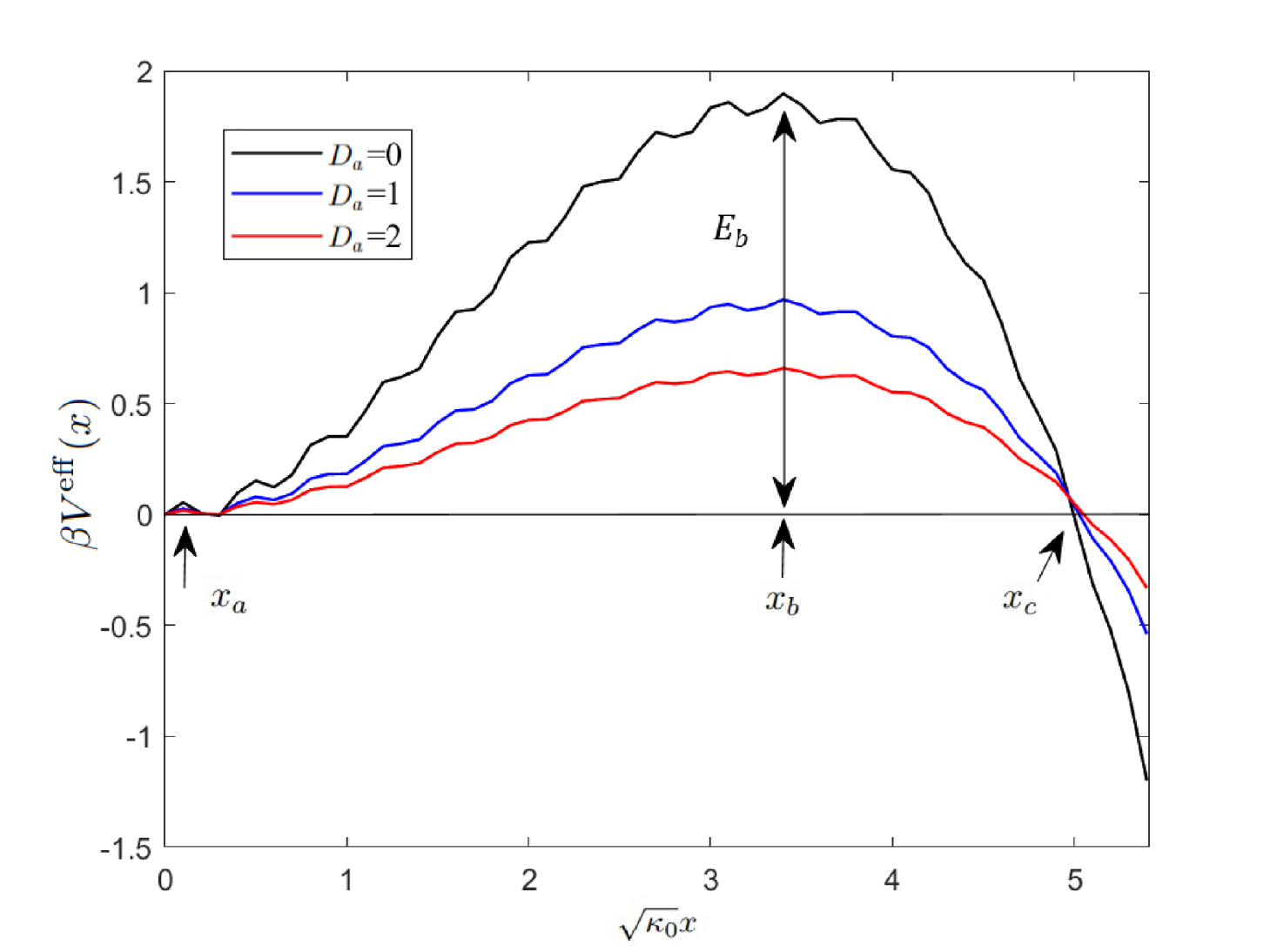}
			\caption{\label{fig1} (Color online) Bare potential and analytic effective potential $\beta V^{{\rm eff}}(x)$, Eq.~(\ref{eq:26}), for different values of $D_a.$ For the given parameter $\alpha\kappa_0^{-3/2}=0.1$, $\tau\kappa_0=0.02$, $\varepsilon=0.05$ and $q\kappa_0^{-1/2}=17$.}
		\end{figure}
		
		From Eq.~(\ref{eq:111}), we obtain   
		\begin{equation}\label{eq:112}
		e^{\psi^{\pm }(x)}=I_0\left(\frac{\varepsilon}{1+D_a}\right),
		\end{equation}
		where $I_0$ is the modified Bessel function$^{[36]}$. Substituting Eq.~(\ref{eq:112}) into Eq.~(\ref{eq:241}), we obtain the escape rate 
		\begin{eqnarray}\label{eq:28}
		r_{\rm act}^{\rm rou}
		=r_{\rm pass}e^{\frac{D_a\left(\beta E_0-\kappa_0\tau\right)}{1+D_a}}\left[I_0\left(\frac{\varepsilon}{1+D_a}\right)\right]^{-2}.
		\end{eqnarray}
		Since the modified Bessel function is always larger than 1, we have $r_{\rm act}^{\rm rou}<r_{\rm act}=r_{\rm pass}e^{D_a(\beta E_0-\kappa_0\tau)/(1+D_a)}$. That is, the roughness due to oscillating perturbation hinders escaping.
		
		Fig.~\ref{fig2} shows the dependence of $r_{\rm act}^{\rm rou} / r_{\rm pass}$ on activity and roughness. $r_{\rm act}^{\rm rou} / r_{\rm pass}$ increases with the increase of activity, but decreases with the increase of roughness.
		
		\begin{figure}\label{fig2}
			\centering
			\includegraphics[width=7cm]{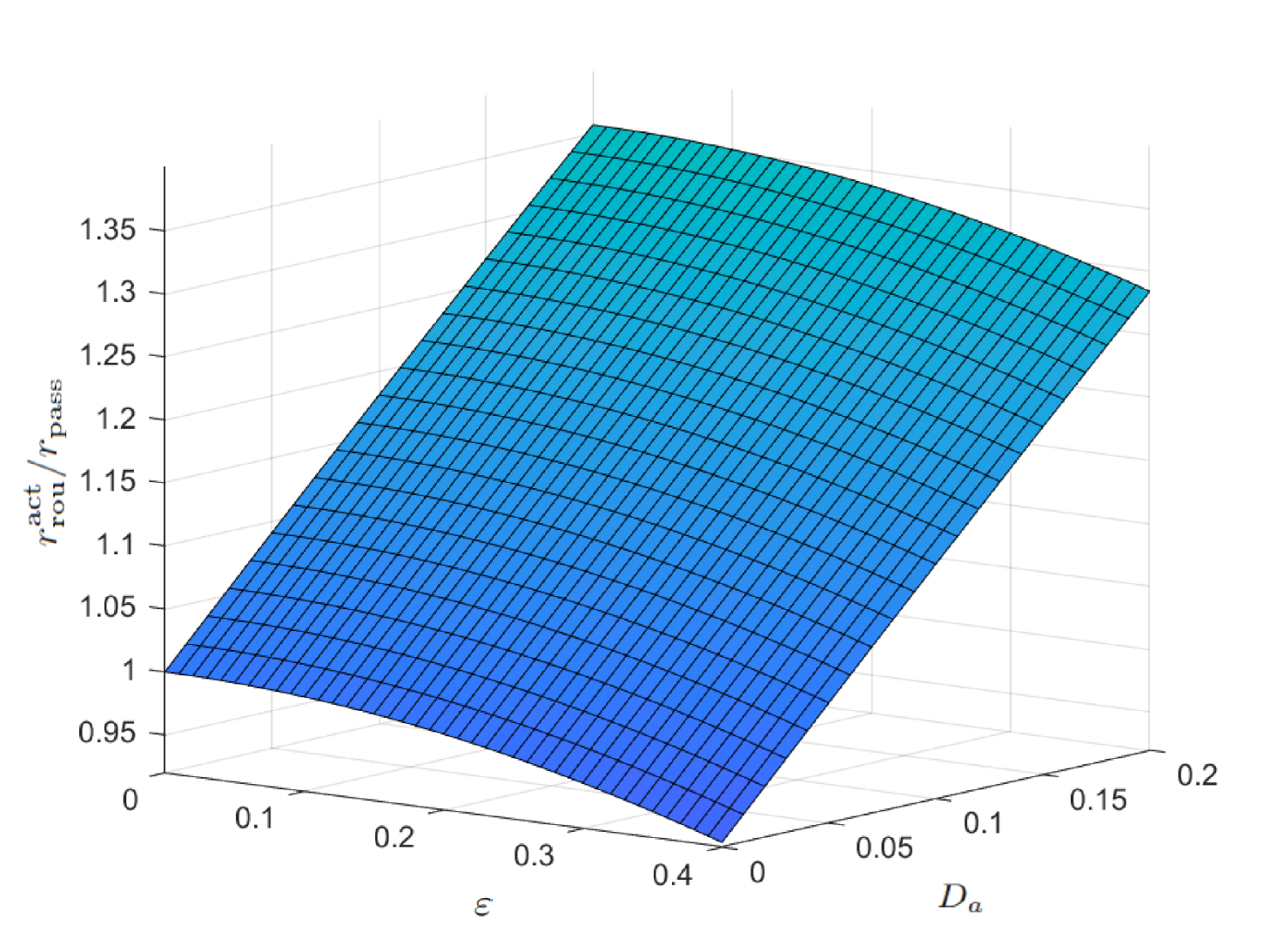}
			\caption{\label{fig2} (Color online) Dependence of $r_{\rm rou}^{\rm act}/r_{\rm pass}$ on amplitude $\varepsilon$ and active parameter $D_a$. Where $\alpha\kappa_0^{-3/2}=0.1$, $\tau\kappa_0=0.02$.}
		\end{figure}

		\subsection{\label{sec:level3B} Random amplitude of rough potential} 
		
		Considering the random amplitude of rough potential $V_1$ with a Gaussian distribution 
		\begin{equation}\label{eq:34}
		\rho(V_1)=\frac{1}{\sqrt{2 \pi\sigma^2}}e^{-\frac{V_1^2}{2\sigma^2}},
		\end{equation}
		where $\sigma$ is standard deviation. The effective rough potential $\beta V^{\rm eff}(x)$ is
		\begin{equation}\label{eq:36}
		\beta V^{{\rm eff}}(x)\approx\frac{1}{2}\kappa_a x^2 -\alpha'x^3+g(x)+ \frac{\varepsilon V_1}{1+D_a}.
		\end{equation}
		From Eq.~(\ref{eq:111}), we obtain 
		\begin{equation}\label{eq:113}
		e^{\psi^{\pm }(x)}=e^{-\frac{\varepsilon^2 \sigma^2}{(1+D_a)^2}}.
		\end{equation} 
		Substituting Eq.~(\ref{eq:113}) into Eq.~(\ref{eq:241}), we obtain the escape rate   
		\begin{eqnarray}\label{eq:38}
		r_{\rm act}^{\rm rou}= r_{\rm pass}e^{\frac{D_a\left(\beta E_0-\omega_0\tau\right)}{1+D_a}}e^{-\frac{\varepsilon^2 \sigma^2}{(1+D_a)^2}}.
		\end{eqnarray}
		Since $e^{-\varepsilon^2 \sigma^2/(1+D_a)^2}$ is always less than 1, we have $r_{\rm act}^{\rm rou}<r_{\rm act}=r_{\rm pass}e^{D_a(\beta E_0-\kappa_0\tau)/(1+D_a)}$. That is, the random amplitude hinders escaping. 
		
		Fig.~\ref{fig3} shows the dependence of $r_{\rm act}^{\rm rou} / r_{\rm pass}$ on activity and roughness. $r_{\rm act}^{\rm rou} / r_{\rm pass}$ increases with the increase of activity, but decreases with the increase of roughness.
		
		\begin{figure}\label{fig3}
			\centering
			\includegraphics[width=7cm]{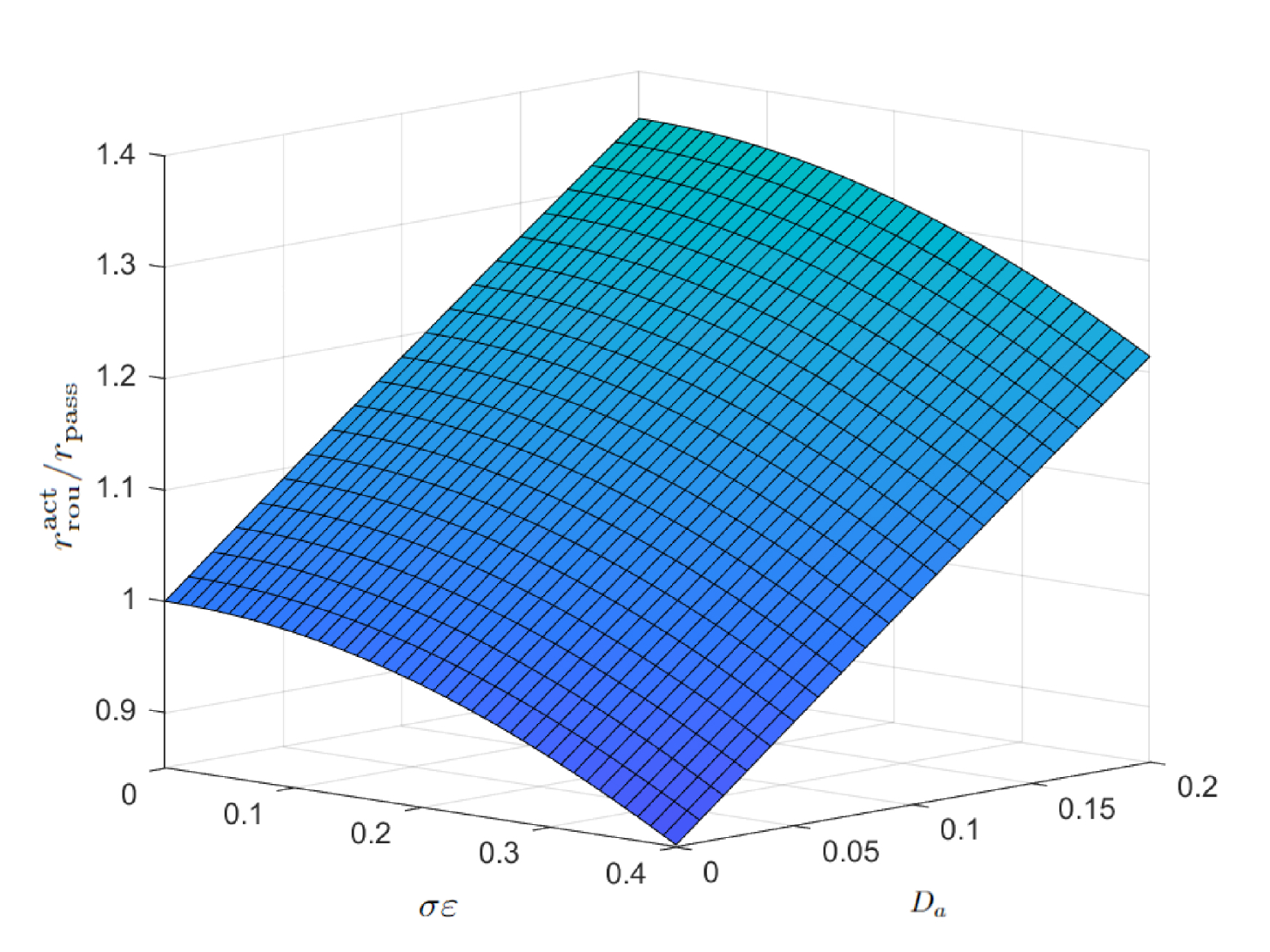}
			\caption{\label{fig3} (Color online) Dependence of $r_{\rm rou}
				^{\rm act}/r_{\rm pass}$ on amplitude $\sigma\varepsilon$ and active parameter $D_a$. Where $\alpha\kappa_0^{-3/2}=0.1$, $\tau\kappa_0=0.02$.}	
		\end{figure} 
		
		\section{\label{sec:level4} Conclusions} 
		
		In this work, we have discussed the escape rate of ABPs in rough potentials by using the effective equilibrium approach and the Zwanzig method. We find that activity usually enhances the escape rate. Both the oscillating perturbation and the random amplitude of rough hinder escaping. In the theoretical derivation, we need the amplitude $\varepsilon$ and $\kappa_0\tau$ are small. Our theory is not appliable for large $\varepsilon$ and $\kappa_0\tau$. We will develop new theoretical approach to deal with these situations in the future.
		
		\section{Acknowledgments} 
		
		The authors are grateful for financial support from the
		National Natural Science Foundation of China (Grant
		No. 11975050 and No. 11735005). We are very grateful for the help of Xiu-Hua Zhao.
		
		\appendix
		
		\section{\label{sec:a} Detailed derivation of Eq.~(\ref{eq:16})}
		
		Following the Kramers method in Ref. $[43]$, we derive the inverse escape rate of ABP in the effective rough potential.
		
		Assume $\beta E_b\gg 1$. In this situation, the system stays the quasi-stationary state such that the probability current $J_{\rm act}^{\rm rou}$ is approximately independent of $x$. By integrating Eq.~(\ref{eq:15}) between $x_a$ and $x_c$ and considering an absorbing boundary condition $x = x_c$, we obtain
		\begin{equation}\label{eq:44}
		J_{\rm act}^{\rm rou}=\frac{D_t e^{\beta V^{\rm eff}(x_a)}\phi (x_a)}{\int_{x_a}^{x_c}\frac{e^{\beta V^{\rm eff}(x)}}{D(x)}dx}.
		\end{equation}
		
		Because the barrier is high, $\phi(x)$ near $x_a$ may be approximately given by the stationary distribution 
		\begin{equation}\label{eq:45}
		\phi(x)=\phi(x_a)e^{-[\beta V^{\rm eff}(x)-\beta V^{\rm eff}(x_a)]}.
		\end{equation}
		The probability $p$ to find ABP near $x_a$ is
		\begin{equation}\label{eq:46}
		p=\int_{x_1}^{x_2}\phi(x)dx=\phi(x_a)e^{\beta V^{\rm eff}(x_a)}\int_{x_1}^{x_2}e^{-\beta V^{\rm eff}}dx,
		\end{equation}
		where $x_1 \leq x_a \leq x_2 \leq x_b$.
		Finally, we can derive Eq.~(\ref{eq:16}) by using $p/J_{\rm act}^{\rm rou}$.
		
		\section{\label{sec:b} Saddle-point approximation}
		
		The integral expression in Eq.~(\ref{eq:201}) may be obtained via the saddle-point approximation at $x_a$ and $x_b$, respectively. 
		
		The effective smooth potential nearly $x_b$ can be expanded nearby $x_b$ as: 
		\begin{equation}\label{eq:49}
		\beta V_0^{\rm eff}(x)=\beta V_0^{\rm eff}(x_b)-\frac{1}{2}\kappa_b(x-x_b)^2.
		\end{equation}
		
		The second integral of smooth potential on the right-hand side of Eq.~(\ref{eq:201}) is expressed as 
		\begin{equation}\label{eq:50}
		\int_{x_a}^{x_c}dx\frac{e^{\beta V_0^{\rm eff}(x)}}{D(x)}
		=\int_{x_a}^{x_c}dx\frac{1}{D(x)}e^{\beta V_0^{\rm eff}(x_b)-\frac{1}{2} \kappa_b(x-x_b)^2}.
		\end{equation}
		According to the spirit of saddle-point approximation, Eq.~(\ref{eq:50}) is transformed into
		\begin{eqnarray}\label{eq:61}
		\int_{x_a}^{x_c}dx\frac{1}{D(x)}e^{\beta V_0^{\rm eff}(x_b)-\frac{1}{2} \kappa_b(x-x_b)^2}
		=\sqrt{\frac{2\pi}{|\kappa_b|}}e^{\beta V_0^{\rm eff}(x_b)}\frac{1}{D(x_b)}.
		\end{eqnarray}
		Substituting $x_b$ into Eq.~(\ref{eq:8b}) and considering $\kappa_0\tau$ is small, we obtain
		\begin{equation}\label{62}
		\int_{x_a}^{x_c}dx\frac{e^{\beta V_0^{\rm eff}(x)}}{D(x)}=\sqrt{\frac{2\pi}{|\kappa_b|}}\frac{e^{\beta V_0^{\rm eff}(x_b)}e^{-{\frac{D_a\kappa_0\tau}{1+D_a}}}}{\left(1+D_a\right)}.
		\end{equation}
		
		Similarly, the first integral of smooth potential on the right-hand side of Eq.~(\ref{eq:201}) may also be obtained by a saddle-point approximation at $x_a$.

	\end{CJK}
\end{document}